\documentclass[aps,prb,twocolumn,showpacs]{revtex4}

\usepackage{graphics}
\usepackage{amstext,amsopn,amsmath,amssymb}
\usepackage{epsfig}

\begin{document}

\title{Ordering kinetics in an fcc $A_3B$ binary alloy model: 
Monte Carlo studies.}

\author{M. Kessler,$^1$ W. Dieterich,$^1$ and A. Majhofer$^2$} 

\affiliation{$^1$Fachbereich Physik, Universit\"at Konstanz D--78457 Konstanz, Germany}
\affiliation{$^2$Institute of Experimental Physics, Warsaw University ul. Ho\.za 
69, PL--00681 Warszawa, Poland}

\date{24 July 2002}

\begin{abstract}                

Using an atom--vacancy exchange algorithm, we 
investigate the kinetics of the order--disorder transition in
an fcc $A_3B$ binary alloy model following a temperature quench 
from the disordered phase. We observe two clearly distinct
ordering scenarios depending on whether
the final temperature $T_f$ falls above or below the 
ordering spinodal $T_{sp}$, which is deduced from 
simulations at equilibrium. For shallow quenches ($T_f>T_{sp}$)
we identify an incubation time $\tau_{inc}$ which characterizes
the onset of ordering through the formation of overcritical
ordered nuclei. The algorithm we use together with experimental
information on tracer diffusion in Cu$_3$Au alloys
allows us to estimate the
physical time scale connected with $\tau_{inc}$ in that 
material. Deep quenches, $T_f<T_{sp}$,
result in spinodal ordering. Coarsening processes at long
times proceed substantially slower than predicted by the 
Lifshitz--Allen--Cahn $t^{1/2}$ law. Structure factors related
to the geometry of the two types of domain walls that appear in
our model are found to be consistent with Porod's law in
one and two dimensions.

\end{abstract}


\maketitle

\section{Introduction}

Several face--centered cubic binary metallic alloys, like
Cu$_3$Au, Cu$_3$Pd, Mg$_3$In, Co$_3$Pt, etc. exhibit long range
order with a $L1_2$--structure below some specific ordering 
temperature $T_0$. In the $L1_2$--structure four equivalent 
phases exist, where one of the four simple cubic sublattices
building the fcc--lattice are preferentially occupied by the 
minority atoms. In studying kinetic processes of phase ordering,
the following general considerations must be taken into account:

\begin{enumerate}  

\item[({\it i})]
The transition is of first order. Hence, for small enough
supercoolings the disordered phase remains metastable. Relaxation 
at short times after a temperature quench is governed by the 
formation of ordered nuclei, which grow into the disordered 
matrix. On the other hand, the concept of spinodal ordering has been 
advanced \cite{AllenCsp} to characterize the phase ordering processes 
under large supercoolings.

\item[({\it ii})]
The degeneracy of the ordered phase implies a four--component 
order parameter, $\Psi = (\psi_0,\psi_1,\psi_2,\psi_3)$. Here, 
$\psi_{\alpha}$ with $\alpha = 1,2,3$, are non--conserved 
structural order parameter components coupled to a conserved
density, $\psi_0$, which describes the concentration of the 
two atomic species. 

\item[({\it iii})] 
The antiphase domain structure is anisotropic as a result of 
the existence of two types of antiphase domain walls:  
low--energy (type--I) and high--energy (type--II) walls. 

\end{enumerate}

Under these conditions a rich spectrum of kinetic processes 
is to be expected. In particular, there remain still open 
questions concerning scaling and universality in the 
late--stage growth kinetics. This has motivated several 
experimental investigations of ordering after a temperature 
quench, mostly on Cu$_3$Au, where direct information has been
obtained from time--resolved X--ray diffraction and neutron 
scattering. On the other hand, only few theoretical or computer 
simulation studies on such materials have been reported so far 
.\cite{Lai,Fron,paper1} Frontera {\it et al.} \cite{Fron} have 
recently simulated the growth of $L1_2$--ordered domains both 
within an atom--atom exchange and the more realistic 
atom--vacancy exchange mechanisms, for quench temperatures
below the expected spinodal temperature $T_{sp}$. A similar 
model with atom--vacancy exchange ($ABV$--model) was applied by 
Kessler {\it et al.} \cite{paper1} to surface--induced kinetic 
ordering processes near Cu$_3$Au(001). 

In this paper we investigate the ordering kinetics  in the bulk 
of such alloys at quench temperatures $T_f$ both below and above 
$T_{sp}$. In contrast to earlier Monte Carlo studies of  nucleation 
in Ising--type models \cite{BindStauf,StaufferA} we again employ the 
$ABV$--model with effective chemical interactions between  nearest 
neighbors  on the fcc--lattice. This allows us to compare our results 
directly  with experiments on fcc--alloys, where vacancy--driven 
processes prevail. In particular, we analyze the nucleation regime, 
where we find a well defined incubation time that sensitively  depends 
on the depth of the quench, $T_0 - T_f$.  Moreover, for $T_f < T_{sp}$, 
with $T_{sp}$ deduced from  static correlations, we observe domain 
patterns typical to  spinodal ordering. In the long--time coarsening
regime we extract  anisotropic scaling functions. Our model is a 
limiting case where type-I domain walls have exactly zero 
formation--energy and therefore are extremely stable. Within our 
accessible time window we find coarsening exponents which are 
significantly smaller  than the conventional exponent $\alpha=1/2$ 
for curvature--driven  coarsening.\cite{Lifsh,AllenC}

After shortly explaining our model and simulation techniques 
in section~\ref{MODEL},  we compare in section~\ref{THERMALACT} 
ordering processes that occur 
for shallow ($T_0 >T_f>T_{sp}$) and deep ($T_f<T_{sp}$) quenches.
In section~\ref{INCUBATION} we define the incubation time 
$\tau_{inc}$ and investigate its dependence on the depth of 
the quench. Anisotropic scaling functions are discussed in 
section~\ref{COARSENING}, while section~\ref{SUMMARY} contains 
a short summary of our results.

\section{Model and simulation method \label{MODEL}}  

We consider a three-dimensional lattice of $L \times L \times L$
fcc--cells with cubic lattice constant $a$ and periodic boundary
conditions in all directions. Unless otherwise stated, we use 
$L=128$. Each site, $i$, of the lattice is occupied either by an 
atom of type $A$, an atom of type $B$, or a vacancy, with an obvious
condition that the corresponding occupation numbers fulfill
$c^A_i+c^B_i+c^V_i = 1$. In accord with the stoichiometry of the
$A_3B$ alloys there are exactly three times as many $A$ atoms
present as $B$ atoms. The number of vacancies is chosen small
enough so that they do not affect static
properties of the system.

In our simplified model \cite{paper1} only nearest neighbor 
atom--atom interactions are taken into account. The corresponding
Hamiltonian then reads:

\begin{equation}
H= \sum_{\langle i,j \rangle}[ V_{AA}c^A_ic^A_j + 
V_{BB}c^B_ic^B_j + V_{AB}(c^A_ic^B_j + c^B_ic^A_j)] ,
\end{equation}

where the sum is restricted to nearest--neighbor pairs.
We are interested here in the transition to the ordered
$L1_2$--structure in which minority atoms ($B$) 
predominantly occupy one of the simple--cubic sublattices of
the original fcc--lattice. It is then natural to assume $V_{BB}>0$,
$V_{AA}<0$ and $V_{AB}<0$. The ground state of the $L1_2$ phase is
fourfold degenerate and corresponds to all $B$--atoms
segregated to exactly one of the simple cubic sublattices. For 
a stoichiometric $A_3B$-alloy the transition occurs at 
a temperature $T_0 \simeq 1.83|J|/k_B$, with 
$J=-\frac{1}{4}(V_{AA} + V_{BB} - 2V_{AB})$ as discussed 
in Ref.~\onlinecite{paper1}  (cf. also Refs. 
\onlinecite{Bind1,SchwL,Fron}). The ordered phase is characterized by 
one conserved, scalar order parameter, $\psi_0$, related to the 
composition and three non--conserved order parameter components 
$(\psi_1, \psi_2, \psi_3)$. These are defined by the following 
equations:

\begin{eqnarray}
\psi_0 = m_1 + m_2 + m_3 + m_4, \nonumber \\
\psi_1 = m_1 - m_2 - m_3 + m_4, \nonumber \\
\psi_2 = m_1 - m_2 + m_3 - m_4,  \\
\psi_3 = m_1 + m_2 - m_3 - m_4,  \nonumber
\end{eqnarray}

where $m_{\alpha}$ are the differences of the mean
$A$-- and $B$--occupation numbers, 
$m_\alpha = \langle{c^A_i}\rangle - \langle{c^B_i}\rangle$ 
for $i \in \alpha$. The index $\alpha = 1\dots4$ enumerates 
the four equivalent simple--cubic sublattices of the
fcc--structure. For a homogeneous $A_3B$ alloy, $\psi_0 = 2$. In
the disordered phase, $\psi_1=\psi_2=\psi_3=0$, while 
the four equivalent ordered phases are described as $(\psi_1,
\psi_2,\psi_3)/\bar \psi$ = $(-1,1,1)$; $(1,-1,1)$; $(1,1,-1)$
and $(-1,-1,-1)$, with a temperature--dependent $\bar \psi$ 
($\bar \psi = 2$ at $T=0$).  For a non--homogeneous system 
Eq.~(2) gives the corresponding local order parameters for each 
(cubic) elementary fcc--cell in terms of weighted averages 
of the occupation numbers computed of all sites belonging 
to the cell. (The weights are taken as the inverse of the numbers 
of neighboring fcc--cells that ``share'' the site in question, 
i.~e.  functions 1/2 and 1/8 are taken for the ``face'' and ``corner'' 
sites, respectively).

As can be seen from Eq.~(2), each of the non--conserved components 
of the order parameter describes a modulation of the $B$--atom 
concentration along one of the cubic axes. For example,  
$\psi_1 \ne 0$ means that the system contains alternating 
$B$--enhanced and $B$--depleted atomic layers along the $x$-axes.
As a consequence of such layerwise arrangement of $B$--atoms, the
(100) peak shows up in an X--ray diffraction experiment in
addition to the (200), (020), (002) and (111) peaks 
characteristic of the underlying fcc--lattice. Similarly,
 $\psi_2$ and $\psi_3$--type ordering leads to additional (010) 
and (001) peaks, respectively.\cite{Warren} For a distance $\vec{k}$ 
from these superstructure peaks the scattering intensity is 
described by the structure factors,

\begin{equation}
S_{\alpha}(\vec{k},t) = \langle{|\Psi_{\alpha}(\vec{k},t)|^2}\rangle,
\end{equation}

which in a non--equilibrium state depend on time $t$. In
Eq.~(3), $\Psi_{\alpha}(k_x,k_y,k_z,t)$ denotes the Fourier 
transform of the order parameter $\psi_{\alpha}(x,y,z,t)$, 
and $\alpha=1,2,3$. The widths of these intensity profiles
are determined by the sizes of antiphase domains.\cite{Warren} 
As mentioned in the Introduction, this system displays two types of
antiphase boundaries, low--energy type--I boundaries with no change
in the arrangement of nearest neighbors, and high--energy type--II
boundaries.\cite{KikuchiC} Since the Hamiltonian, Eq.~(1), involves
only nearest neighbor interactions, type--I boundaries in fact
have zero energy.\cite{degeneracy}   
 
We assume here that the time--evolution of our model occurs
only through atom--vacancy exchange processes. In an elementary
move we first randomly single out one of the vacancies and,
second, we choose at random one of its nearest neighbors occupied 
by an atom. The standard Metropolis algorithm is then used to
decide whether the atom--vacancy exchange for the chosen pair
takes place or not. In one Monte Carlo step (MCS), the system
completes a series of as many such elementary moves as there are
lattice sites in the system, i. e.  $4L^3$, so that this time unit
does not depend on the actual number of vacancies. 

In previous simulations \cite{paper1} we have chosen the interaction
parameters such that they were consistent with the observed
Au--segregation at the Cu$_3$Au (001)--surface. Following 
Ref.~\onlinecite{paper1},
we take $V_{BB} = -V_{AA} = -V_{AB} > 0 $ and assume 
$c^V \simeq 6.1 \cdot 10^{-6}$ as the mean 
density of vacancies present in the system.\cite{vacancy} For vacancy
concentrations of that order we have verified that both static
and kinetic properties are independent of the precise value
of $c^V$. To simulate a sudden quench from high temperatures to
a final temperature $T_f<T_0$ we start with a random distribution
of atoms at time $t=0$ and let the system evolve in time at $T_f$.
The ensuing equilibration process is analyzed by calculating
energies, structure factors, and other ordering characteristics
from averages over 10 independent Monte Carlo runs.

\begin{figure}
\epsfig{file=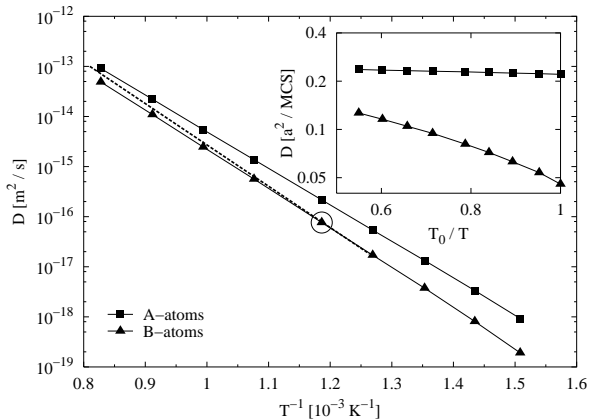, width=8cm}
\caption{Tracer diffusion coefficients of $A$ (full squares)
and $B$ (full triangles) atoms in an $A_3B$ fcc--alloy. Physical 
units were assigned to the Monte Carlo 
data with the help of the experimentally known order--disorder 
transition temperature, $T_0 = 663$~K, and temperature--dependent 
concentrations of vacancies\cite{vacancy} in Cu$_3$Au. The dashed line 
represents measured Au tracer diffusion coefficients.\cite{diffusion1} 
The open circle marks the experimental point used 
to set the time scale. In the inset the corresponding  
``raw'' Monte Carlo results are displayed.}
\end{figure}

As the first application of our atom--vacancy exchange algorithm
we have calculated tracer diffusion coefficients for $A$-- and
$B$--atoms in the disordered phase close to $T_0$. Since in our model
$V_{AA}=V_{AB}$, the jump of an $A$--atom will leave the interaction
energy with its environment almost unchanged and hence $D_A$ is found 
to be nearly temperature--independent. The $BB$--repulsion, on the other
hand, gives rise to a temperature-dependent $D_B<D_A$, see the inset of
Fig.~1. As these results refer to a fixed vacancy concentration,
they cannot be directly related to experiments, where $c^V(T)$ normally 
is a strong function of $T$.\cite{vacancy} Nevertheless, the ratio
$D_A/D_B \simeq 2$ at $T/T_0=1.7$ very roughly agrees with the 
experimental value $D_{Cu}/D_{Au} \simeq 1.45$ for Cu$_3$Au at the
same $T/T_0$--ratio,\cite{diffusion} thus supporting our choice of
nearest--neighbor interaction parameters.
Moreover, in an attempt to map the Monte Carlo time to the physical time
scale, we can exploit experimental knowledge of $c^V(T)$ for 
Cu$_3$Au.\cite{vacancy} As $c^V \ll 1$, hopping rates of a tracer atom are 
simply proportional to $c^V$. This suggests to introduce a new Monte Carlo
time unit 1~MCS$^*=c^V(T)\cdot 1$~MCS, where 1~MCS was defined above as 
one attempted vacancy exchange per lattice site. 
Regarding the simulated mean--square displacements as a function
of the new time scale (with units MCS$^*$) allows us to extract diffusion
constants $D_A$ and $D_B$ as shown in Fig.~1. Scales on the axes
were obtained using the Cu$_3$Au transition temperature $T_0=663$K, the
lattice parameter $a=3.74$\AA\ and an additional parameter, which is 
determined by equating the calculated diffusion constant $D_B$ 
 with the experimental $D_{Au}$ at $T/T_0=1.27$.\cite{diffusion1} The 
last parameter converts 1~MCS$^*$ directly into seconds. The favorable 
comparison between calculated and measured diffusion constants suggests 
that the above procedure, based on the atom--vacancy exchange algorithm 
and the knowledge of the experimental $c^V(T)$--function, provides a 
description of processes on the physical time scale. We shall come back 
to this issue in section~\ref{THERMALACT} in the context of nucleation 
processes.

\section{Thermally activated versus continuous ordering \label{THERMALACT}}

At $T=T_0$ the model--system defined in the preceding section
undergoes a first order phase transition
\cite{Bind1,SchwL,Fron,paper1} from the disordered phase
(for $T>T_0$) to the $L1_2$--type ordered structure (for
$T<T_0$). The correlation length of the disordered phase, $\xi$,
remains finite at coexistence, and approximately fulfills the
mean--field type relation $\xi^2(T) \propto 1/(T-T_{sp})$, with
$T_{sp}<T_0$. In our previous work \cite{paper1} we found
$T_{sp}= (0.967 \pm 0.003) T_0$ as the value best fitting our Monte Carlo 
data for the correlation length above $T_0$. Hence, within this 
procedure, our model displays a fairly well-defined temperature for the
onset of spinodal ordering. A precise separation
of the metastable (thermally activated ordering) 
from the unstable regime (continuous ordering), however, 
does not exist in systems with short--range interactions.\cite{BindStauf,Bind2}

Below we demonstrate the role of that temperature
deduced from equilibrium considerations, in non--equilibrium
ordering processes. It turns out that for temperatures
$T_f>T_{sp}$ and $T_f<T_{sp}$ two contrasting transition scenarios
occur that lead to domain patterns typical of nucleation and
spinodal ordering, respectively. We shall illustrate this
by one example for each regime.

Let us start our description with thermally activated growth of
the ordered phase, i.~e. with the case of a shallow quench,
$T_0>T_f>T_{sp}$. We then expect that the ordered regions -- nuclei 
of the new phase -- are repeatedly formed and destroyed by thermal
fluctuations unless a nucleus exceeding a certain critical size
is built.  Once formed, such an overcritical nucleus will grow
relatively fast against the surrounding disordered bulk until it
meets another (growing) ordered domain.
An $xy$--section of a configuration with overcritical nuclei formed in
a system of $128 \times 128\times 128$ fcc--cells is shown in
Figs.~2a and 2b. It emerged 7000~MCS after the quench to
$T_f=0.972\,T_0>T_{sp}$. The central nucleus seen in the figure
is in fact composed of three ``twinned'' crystallites,
corresponding to $\psi_1 \simeq 2$ (the large, nearly
homogeneous, bright spot in Fig.~2a) and $\psi_1 \simeq -2$ (the
neighboring dark spot). As shown in Fig.~2b, the homogeneous 
bright spot of Fig.~2a is in fact composed of two crystallites 
corresponding to $\psi_3 \simeq -2$ and $\psi_3 \simeq 2$, 
respectively. A second nucleus characterized by 
$\psi_1 \simeq 2$ and $\psi_3 \simeq -2$ grows at the right 
edge of the system. The configuration displayed in Figure~2 
is quite typical for a quench temperature slightly above $T_{sp}$.

\begin{figure}
\begin{center}
\epsfig{file=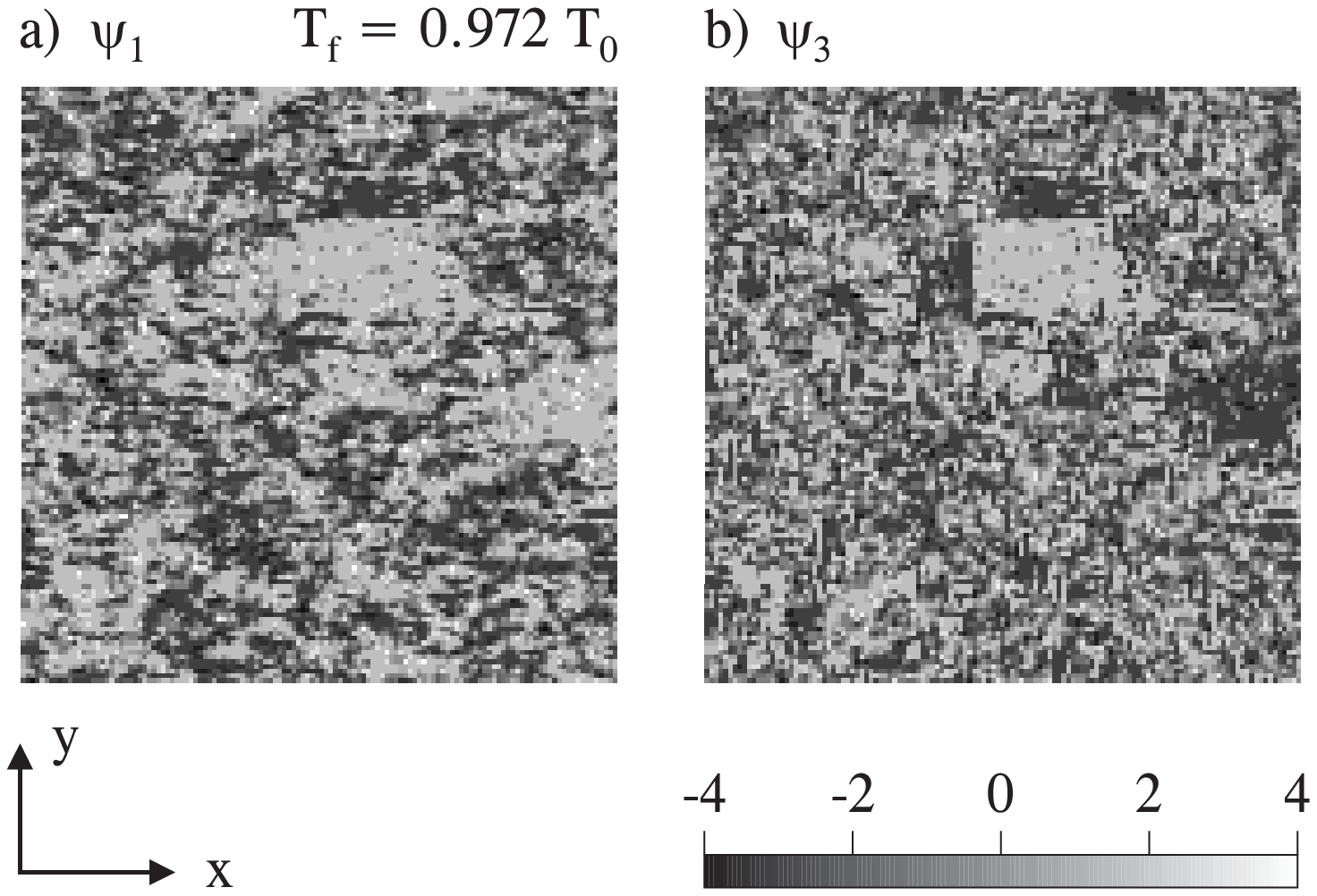, width=7.5cm}
\epsfig{file=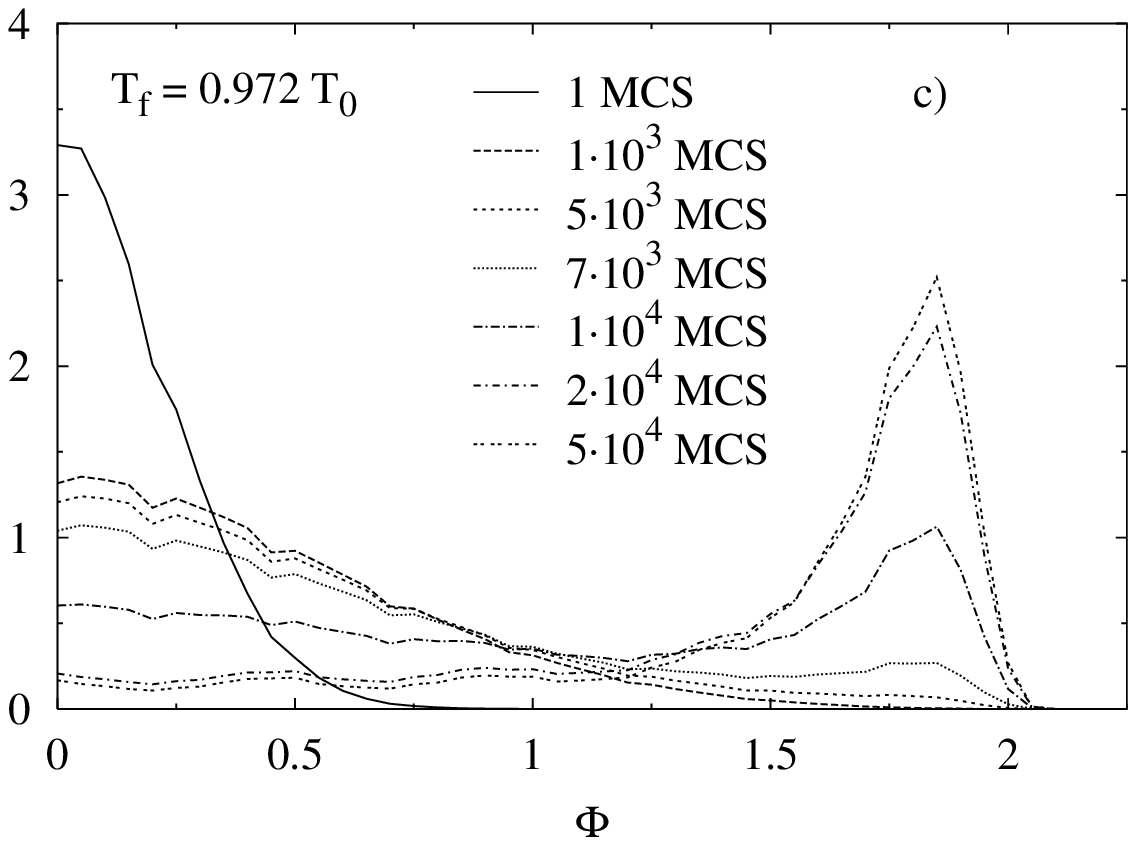, width=8.4cm}
\end{center}
\caption{Thermally activated ordering in a system of 
$128\times 128\times 128$ fcc--cells at $T_f=0.972\,T_0>T_{sp}$.
(a) $\psi_1$  and (b) $\psi_3$ patterns containing overcritical 
nuclei at a time $t=7\times 10^3$~MCS after the quench from random 
initial conditions (grey scales indicate local values of 
$\psi_{\alpha}$ ($\alpha={1,3}$) between $-4$ and $4$). 
(c) Histograms of corresponding  
$\Phi = (|\psi_1|+|\psi_2|+|\psi_3|)/3$ values averaged 
over blocs of $4\times 4\times 4$ fcc--cells calculated 
at a series of times after the quench.}
\end{figure}

To follow the ordering process that takes place in the whole volume
of the system we now introduce an additional quantity \cite{Fron}
$\Phi=(|\psi_1|+|\psi_2|+|\psi_3|)/3$ as a convenient indicator of
the degree of local order:  $\Phi \simeq 0$ in the disordered phase
and $\Phi \le 2$ for any of the four equivalent ordered states.
Further, we divide the system into blocs of $4\times 4\times 4$
fcc-cells, calculate $\Phi$ independently for each bloc to obtain
a distribution function $P(\Phi,t)$ for a series of time intervals 
after the quench. For a quench to $T_f=0.972\,T_0$ the obtained  
histograms are shown in Fig.~2c. In this figure we easily identify 
three stages of the ordering process. First, up to several thousand MCS, 
the whole system remains disordered. The distribution of $\Phi$ broadens 
with time but stays concentrated close to $\Phi=0$. Then, after some
incubation time, $\tau_{inc}$, see section~\ref{INCUBATION}, 
the second maximum close to $\Phi=2$ shows up. At that moment first 
overcritical nuclei emerge in the system and start to grow. The second
stage, in which overcritical nuclei grow against the disordered
bulk, is realized by histograms for $7 \cdot 10^3 \le t\le 10^4$~MCS
in which both peaks are clearly visible. 
In the third stage, practically the whole system is already
ordered and, as illustrated by curves for $t=2 \cdot 10^4$ and
$5 \cdot 10^4$~MCS, the distribution of $\Phi$ changes very little with
time. At that stage coarsening processes take place.

\begin{figure}
\begin{center}
\epsfig{file=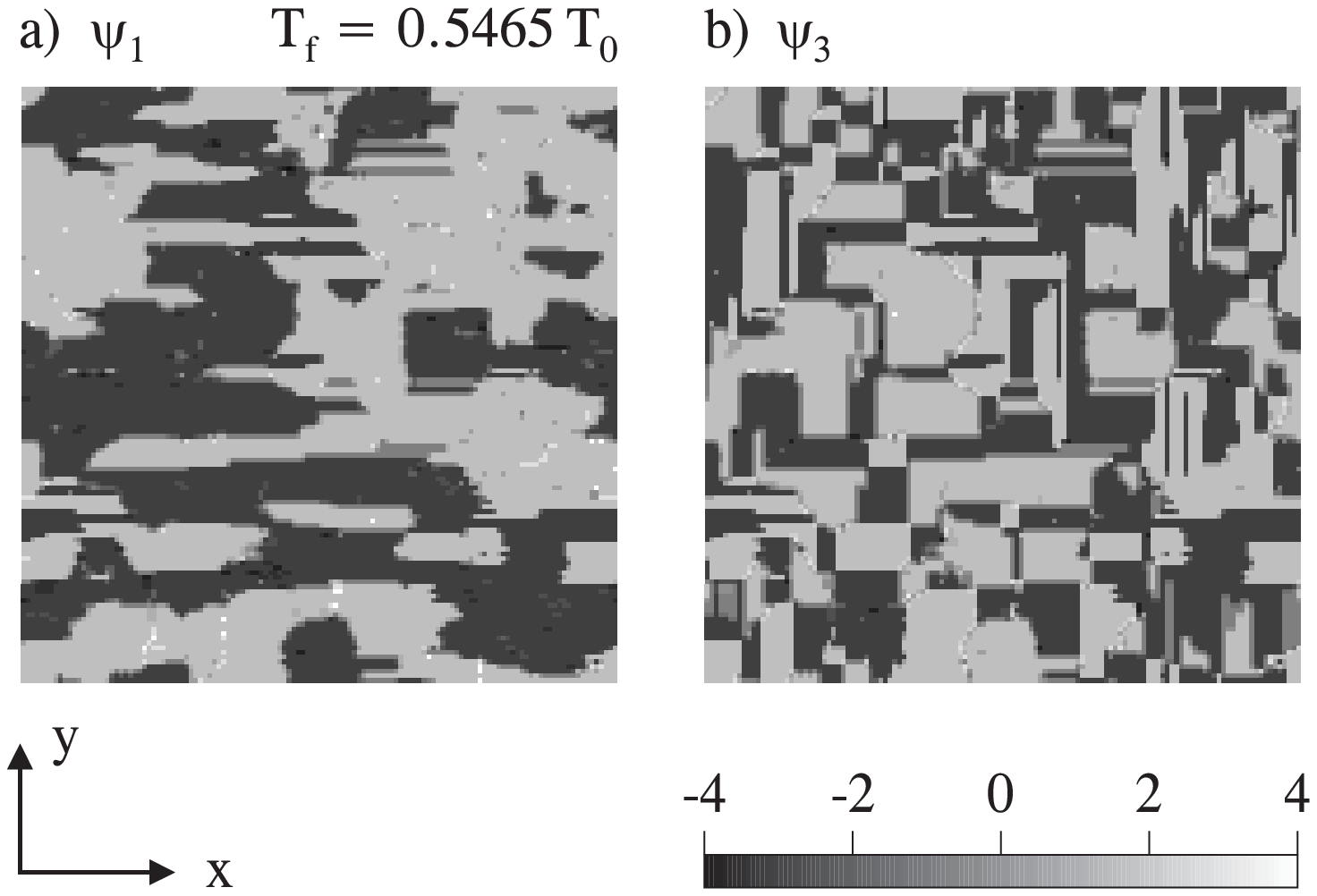, width=7.5cm}
\epsfig{file=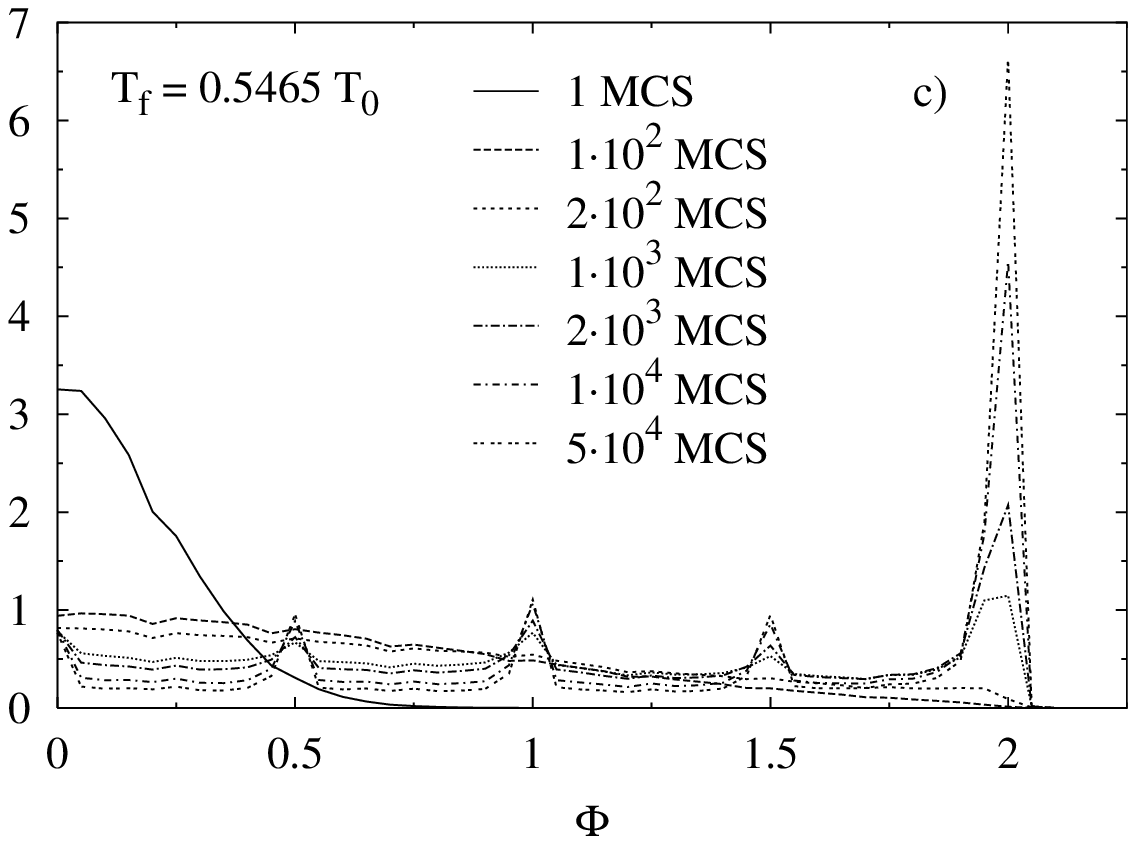, width=8.4cm}
\end{center}
\caption{Continuous ordering in a system of 
$128\times 128\times 128$ fcc--cells at $T_f=0.5465\,T_0<T_{sp}$.
(a) $\psi_1$ and (b) $\psi_3$ patterns at a time 
$t=7\times 10^3$~MCS after the quench from random initial 
conditions. Two types of domain walls are clearly shown.      
(c) Histograms of corresponding $\Phi$ values calculated at 
a series of times after the quench.}
\end{figure}

In contrast to the above sequence of events, deep quenches
($T_f<T_{sp}$) result in continuous ordering, in which shortly
after the quench the whole volume of the system decomposes into
ordered domains. As a representative example Figs.~3a and 3b
display distributions of $\psi_1$ and $\psi_3$, respectively,
for an $xy$--section through the system at $t=7 \cdot 10^3$~MCS after 
a quench to $T_f=0.5465\,T_0$. The two types of domain walls, that may 
be formed in the system,\cite{KikuchiC,Lai} show up in Figs.~3a and 3b. 
The low--energy interfaces appear there as straight lines parallel 
to the $x$ or $y$--axis. Across each of these lines $\psi_1$ and 
$\psi_3$, or $\psi_2$ and $\psi_3$ simultaneously change sign across 
line--intervals parallel, to the $x$ or $y$--axes, respectively.
Within the simple model used in this paper, formation of such walls
costs no energy and therefore they are very stable. The curved
interfaces seen in Fig.~3a consist of sectors of
high--energy domain walls across which $\psi_1$ and $\psi_2$
change sign simultaneously.  The histograms of $\Phi$--values 
obtained after averaging over 10 independent runs for   
$T_f=0.5465\,T_0$ are shown in Fig.~3c for a sequence of time
intervals after the quench.  There, in contrast to Fig.~2c, the
initial peak at $\Phi=0$ spreads out within the first 100~MCS
and the maximum at $\Phi \simeq 2$ starts to grow by 200~MCS to
reach a considerable height already at $2 \cdot 10^3$ ~MCS.  The
$\Phi\simeq 2$ peak is very narrow, indicating a nearly perfect
local order in the system.  Additional small, sharp peaks at
multiples of $\Phi=0.5$ may be traced back to the domain walls
that cross some of the blocs of $4\times 4\times 4$ fcc--cells
used to prepare Fig.~3c. Slow coarsening processes, which 
in this case set in already for $t > 2 \cdot 10^3$~MCS result in 
considerable sharpening of the main $\Phi \simeq 2$ peak.

\begin{figure}
\epsfig{file=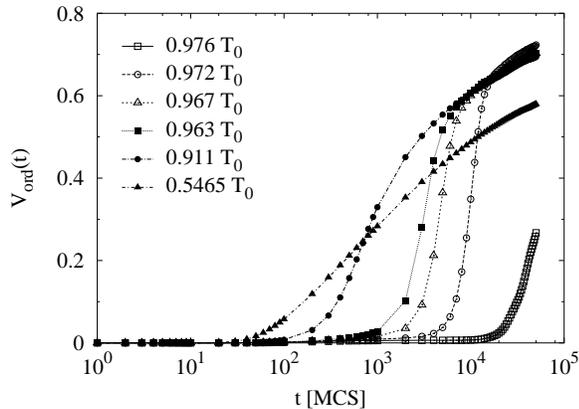, width=8cm}
\caption{Volume fraction, $V_{ord}$, of an ordered phase 
($\Phi>1.5$) as a function of time after the quench for a 
series of temperatures $T_f$.}
\end{figure}

To better illustrate the differences between the two ordering
scenarios described above we show in Fig.~4 how $V_{ord}$, the
volume fraction of the ordered phase, grows with time for a
number of final temperatures. There, we counted as ordered all
the blocs (of $4\times 4\times 4$ fcc--cells each) for which
$\Phi >1.5$. The character of $V_{ord}(t)$ changes from a slowly
increasing function of $t$ for $T_f \le 0.911\,T_0$ to a sharp,
nearly stepwise growth after some incubation time for
$T_f>0.972\,T_0$.

In the next two sections we analyze in more detail the incubation
time and the long time coarsening processes.

\section{Incubation time \label{INCUBATION}}

In the preceding section we introduced the incubation time
$\tau_{inc}$ as the characteristic time interval between the quench 
and the observed growth of ordered domains
in the case of metastability, $T_f>T_{sp}$. Let us now look more
precisely at the processes that take place in the system within
the initial stages of ordering. One relevant quantity here is the 
excess energy, $\Delta E(t) = E(t) - E(\infty)$, defined as the 
difference between the actual energy of the system, $E(t)$, and the 
energy it would reach after complete equilibration, $E(\infty)$. This 
means that we identify $E(\infty)$ with the energy of a single ordered 
domain of the size of the whole system, equilibrated at $T_f$.
In Figure~5a  $\Delta E(t)$ per lattice site is shown
for a number of final temperatures $T_f$. Initially, the
relaxation of $\Delta E(t)$ depends on temperature only weakly. 
Then, after about 20~MCS, the curves in Fig.~5a
split due to a considerable slowing down of the relaxation rate
with growing temperature. In turn, for shallow quenches,
ordering proceeds in three stages, as 
displayed already in Fig.~2c. First, the system remains
disordered up to a certain incubation time $\tau_{inc}$.
The metastability of the ordered phase is reflected by 
plateaus in $\Delta E(t)$, which develop near $T_f\simeq T_{sp}$,
and extend with increasing $T_f$. Second, $\Delta E(t)$ drops
markedly when overcritical nuclei are produced by thermal
fluctuations. We define $\tau_{inc}(T)$ as precisely that moment 
after the quench at which, for a given temperature $T$,  
$\Delta E(t)$ drops notably (by about 10 percent) below its
plateau--value. The ordered nuclei formed within the incubation time 
grow against the disordered bulk until, in a third stage, ordered regions 
meet and slow coarsening processes set in.  

In addition to $\Delta E(t)$, we also show in Figure~5b and 5c 
how the corresponding first moments of the structure factors
$S_{\alpha}(k_x,k_y,k_z,t)$ change with time. To account for the
anisotropy illustrated in the domain patterns 
in Fig.~3a and b, we consider the two kinds of structure 
factors
 
\begin{equation}
S_{\parallel}(k,t) = \frac{1}{3}
                   (S_1(k,0,0,t)+S_2(0,k,0,t)+S_3(0,0,k,t)),
\end{equation}

and

\begin{eqnarray}
S_{\perp}(k,t) &=& \frac{1}{N_k}\sum \limits_{q_1^2 + q_2^2=k^2}
     (S_1(0,q_1,q_2,t) \nonumber\\
   &&  + S_2(q_1,0,q_2,t) + S_3(q_1,q_2,0,t)),
\end{eqnarray}

where $N_k$ denotes here the number of $(q_1,q_2)$ pairs that fulfill
$q_1^2 + q_2^2 = k^2$, as well as their first moments

\begin{equation}
k_{\parallel}(t) = \frac{\sum_{k>0} k S_{\parallel}(k,t)}
                        {\sum_{k>0} S_{\parallel}(k,t)},
\end{equation}

\begin{equation}
k_{\perp}(t) = \frac{\sum_{k>0} k S_{\perp}(k,t)}
                   {\sum_{k>0} S_{\perp}(k,t)}.
\end{equation}

\begin{figure}[ht!]
\begin{center}
\epsfig{file=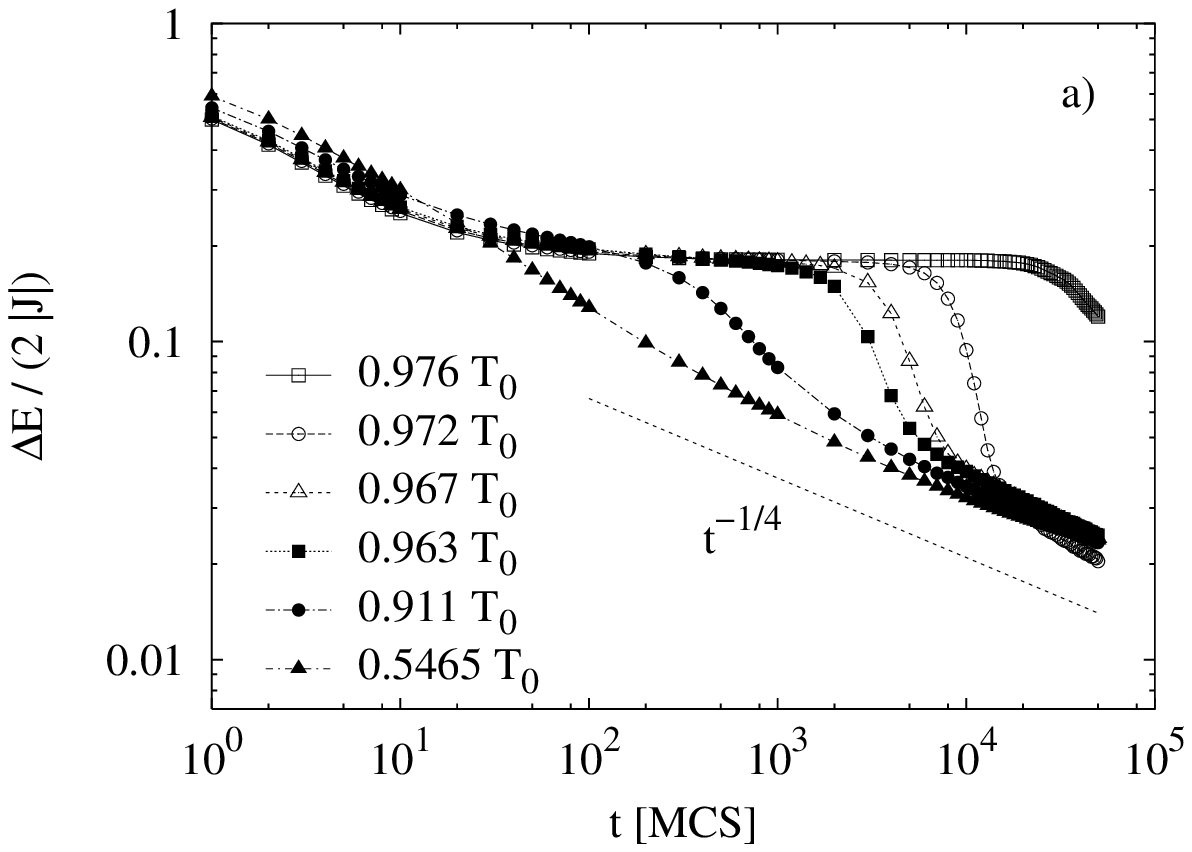, width=8cm}
\epsfig{file=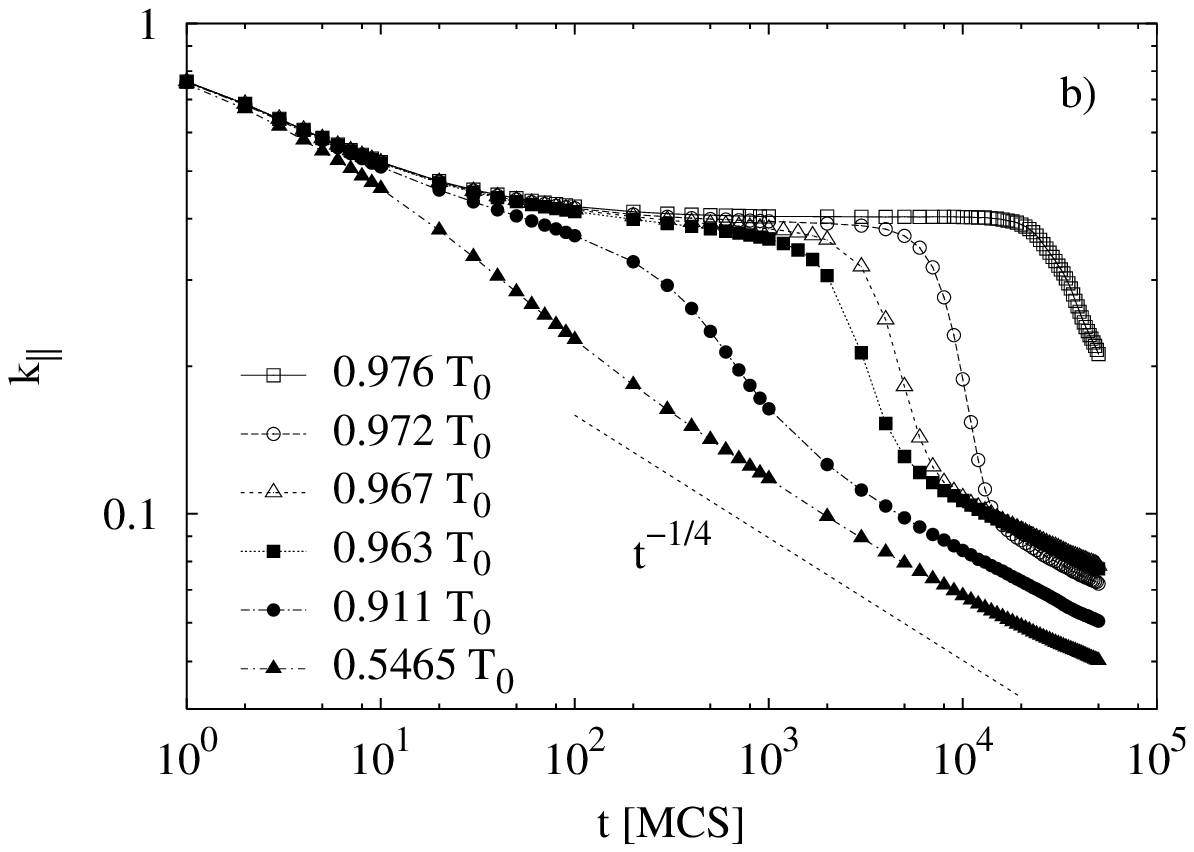, width=8cm}
\epsfig{file=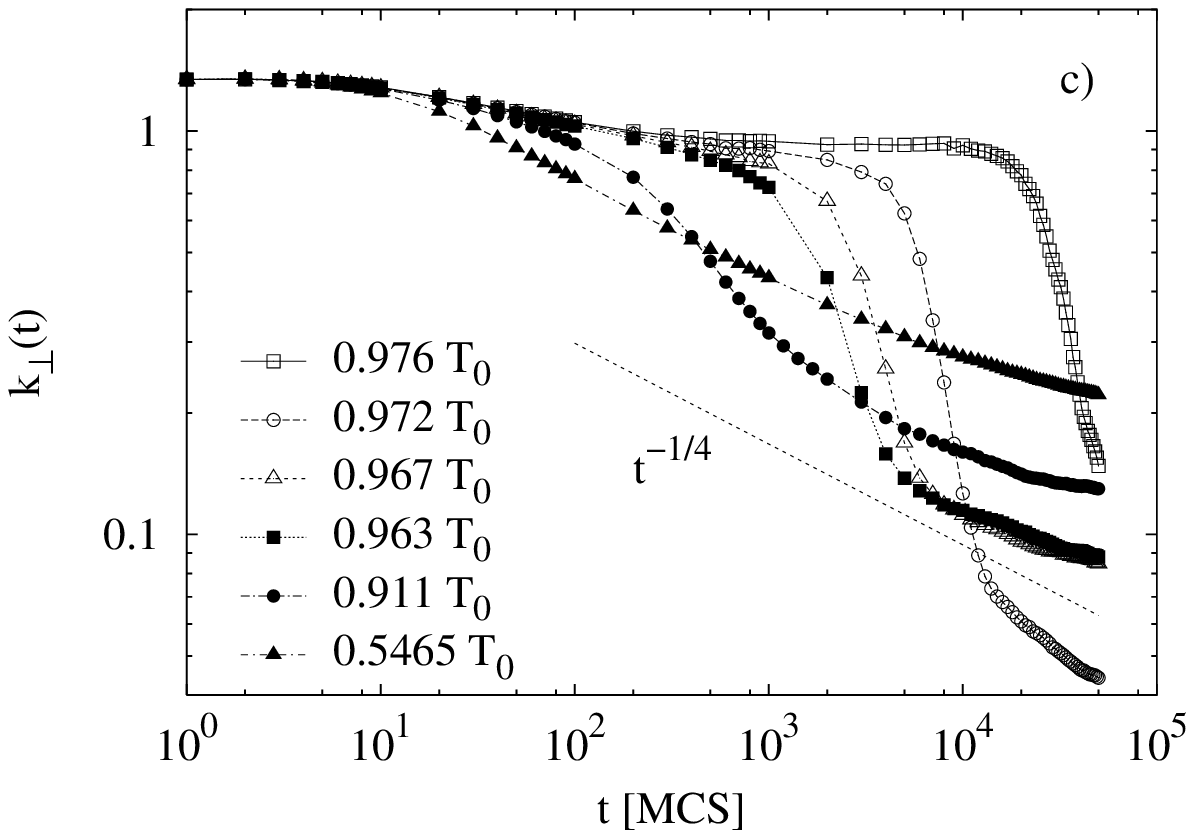, width=8cm}
\end{center}
\caption{Excess energy per lattice site $\Delta E$ (a), and the first moments
$k_{\parallel}(t)$ (b) and $k_{\perp}(t)$ (c) of the structure factors
$S_{\parallel}(k,t)$ and $S_{\perp}(k,t)$ 
as functions of time after the quench,  
calculated for a series of final temperatures $T_f$.}
\end{figure}

In Eq.~(6) $k$ denotes one wave--vector component, see Eq.~(4),
while in Eq.~(7) $k$ is defined as in Eq.~(5).		   
Clearly, $S_{\parallel}(k,t)$ is determined by structural modulations
due to high--energy walls, which are reflected, for example, by 
sign--changes of $\psi_1$ when going along the $x$--axis, see Fig.~2. 
The quantity $k_{\parallel}$ therefore characterizes the inverse 
distance between such walls in one symmetry direction. On the other 
hand, $S_{\perp}$ is sensitive to the two-dimensional network
of low--energy walls, and $k_{\perp}$ gives the inverse distance
between such walls, cf. Fig.~3.

Curves displayed in all three parts of Figure~5 are 
strikingly similar in form, confirming that relaxation processes
are indeed dominated by ordered domains growing in size.
Moreover, $\Delta E$, $k_{\parallel}$ and $k_{\perp}$, all
start to drop below their plateau--values at the same moment
after the quench. This fact corroborates our linking
the plateaus in $\Delta E(t)$ to the temperature-dependent incubation 
time for the formation of overcritical ordered nuclei. In accord with 
the spatial patterns shown in Figs.~2a, 3a and 3b, where high--energy 
walls are much further apart from each other than the low-energy ones, 
$k_{\parallel}(t)$ is in most cases considerably smaller than 
$k_{\perp}(t)$ calculated at the same temperature.

There are two effects restricting the range of $\tau_{inc}(T_f)$
accessible to simulations. First, the system size $L$ must be 
considerably larger than the size of overcritical nuclei which are 
formed at temperature $T_f$. Second, obtaining smooth
curves in Fig.~5 requires a sufficiently large number of 
nucleation events within the maximum computation time $t_{max}$. Again, 
this becomes increasingly 
difficult to fulfill with growing $T_f$ because both $\tau_{inc}$
and the dispersion of nucleation times over the different samples
strongly increase with $T_f$. Comparing our results up to 
$t_{max}=5 \cdot 10^4$ for lattices of size L=96 and L=128 we find 
$T_f \simeq 0.973\,T_0$ to be roughly the highest temperature for 
which $\tau_{inc}$ can be determined in a reliable way.  

\begin{figure}[b!]
\epsfig{file=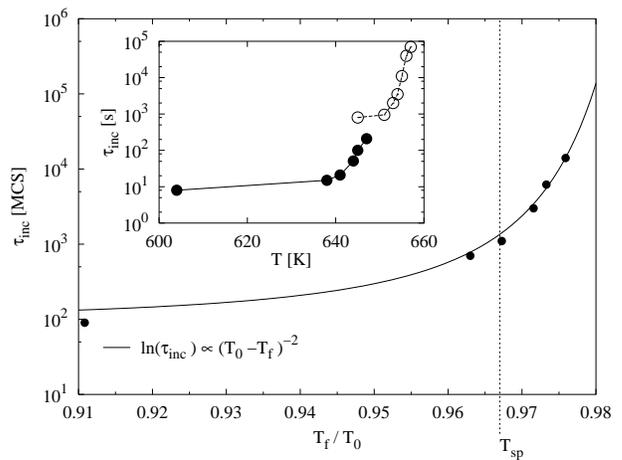, width=8cm}
\caption{Incubation times $\tau_{inc}$ versus reduced
temperature $T_f$. The inset shows the same data (full symbols) after
conversion to the physical time scale as described in the text, together
with incubation times measured \cite{Noda} in Cu$_3$Au (open symbols).}
\end{figure}

Results of our simulations for $\tau_{inc}(T)$, plotted in Fig.~6,
are consistent with the expression

\begin{equation}
\ln\tau_{inc} \propto (T_0-T_f)^{-2}, 
\end{equation}

which is similar in form to the nucleation rates obtained
within classical nucleation theory \cite{Zettlemoyer,BindStauf} 
and Monte Carlo simulations.\cite{BindStauf,StaufferA}
Note that the data point with $T_f/T_0=0.911$ falls well below the spinodal,
but in Fig.~2 there still exists a shoulder at that temperature
as a remnant of the plateaus at higher temperatures.

Incubation times for Cu$_3$Au have been measured by Noda et al.
\cite{Noda} from the width $\Gamma\propto k_{\perp}(t)$ of the (110)
X-ray diffraction peak. Close to $T_0$ the width as a function of time
develops plateaus qualitatively similar to Fig.~5c. Their data, however,
refer to $T_f/T_0>0.981$, whereas ours are for $T_f/T_0\lesssim 0.976$.
Also one should note that experimentally the ratio $T_{sp}/T_0$ may
be sample--dependent. Because of the sensitivity of $\tau_{inc}$ to
temperature differences with respect to $T_{sp}$, this introduces 
considerable uncertainties in our comparison. Nevertheless it is 
interesting to present experiments and simulations in one plot, as
done in the inset of Fig.~6. There the simulation data were linked
to the physical time scale exactly in the manner described towards the
end of section~\ref{MODEL}. Both sets of data display a remarkable 
similarity with respect to the order of magnitude in time and their 
trend with temperature  (and even seem to be a continuation of each other). 
We conclude that the connection between 
time scales for diffusion (see section~\ref{MODEL}) and nucleation as 
implied by our algorithm  roughly agrees with experiment.

\section{Coarsening regime and scaling \label{COARSENING}}

As pointed out in the preceding sections, the distances between 
high-- and low--energy walls define two different length scales. 
Moreover, the high--energy walls that are much further apart from 
each other than the low--energy ones, are also less stable.  As a 
consequence, we expect direction--dependent scaling laws to hold for 
the structure factors $S_{\alpha}(\vec{k},t)$. 
This is made explicit by considering the functions $S_{\parallel}$
and $S_{\perp}$ introduced in Eqs.~4 and 5. Due to their definitions
we expect these structure factors to obey scaling laws for 
one--dimensional and two--dimensional systems, respectively.
This is verified by our simulations
at $T_f=0.5465\,T_0$, i.~e. in the case of continuous ordering. As shown 
by the $1d$ and $2d$ scaling plots in Figs.~6a and 6b, the data taken
for a number of times $t\ge 500$~MCS after the quench indeed collapse
onto single master curves.
Moreover, the decay of both quantities at large
k appears to be consistent with Porod's law in
$d=1$ and $d=2$, respectively,\cite{Porod,comF} 

\begin{equation}
k_{\parallel}(t)S_{\parallel}(k,t) 
   \propto (k/k_{\parallel}(t))^{-2},
\end{equation}

\begin{equation}
k_{\perp}^2(t)S_{\perp}(k,t) \propto (k/k_{\perp}(t))^{-3} . 
\end{equation}

We now ask whether the same scaling laws apply to the    
results for $T_f>T_{sp}$, i.~e. to the case of thermally 
activated ordering.  Figure~8 contains the corresponding 
plots of $S_{\parallel}(k,t)$, part (a), and $S_{\perp}(k,t)$, 
part (b), calculated for $T_f=0.972\,T_0$ at times 
$t>\tau_{inc}$. In both parts of the figure
scaling regimes are observed which are 
consistent with Eqs.~(9) and (10) in an intermediate  
range of $k$--values. Both $S_{\parallel}$ and $S_{\perp}$, however,
develop tails at large $k$ which show significant deviations 
from scaling. 
These deviations can be traced back to irregular shapes of 
overcritical nuclei growing against the disordered bulk 
(cf.~Fig.~2). Such highly structured grain boundaries 
eventually meet, leading to likewise structured domain walls 
when the regime of slow coarsening is reached. Small inclusions of 
disordered phase still remain trapped between boundaries of ordered 
grains for quite a long time after the onset of coarsening processes 
in the bulk of the crystal.

Closer inspection of our data near the onset of those 
tails reveals that indeed, with increasing time $t$, the
validity of Porod's law extends to larger $k$ in both parts 
of Figure~8. We may take this as indication that 
Porod's law, defined by Eqs.~(9) and (10), will dominate in the 
late stages of continuous as well as thermally activated ordering 
processes. 

\begin{figure}
\epsfig{file=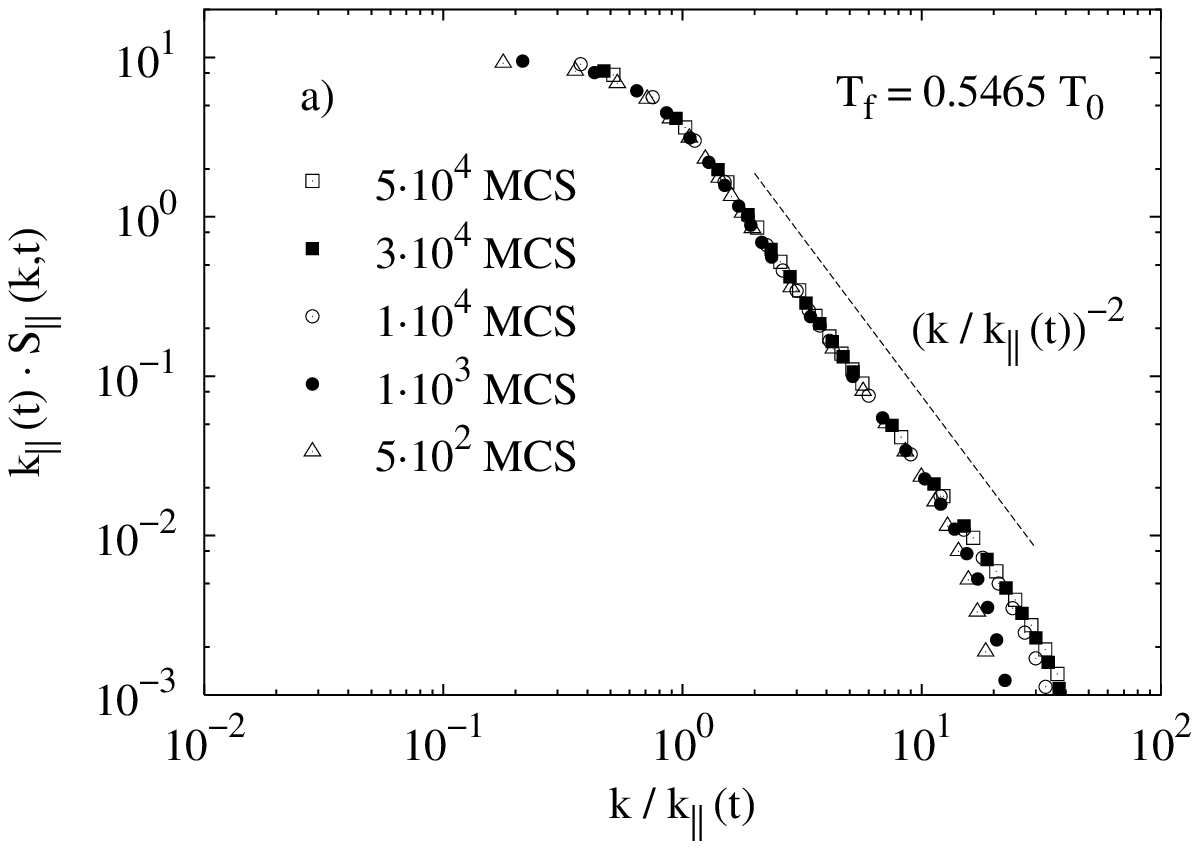, width=8cm}
\epsfig{file=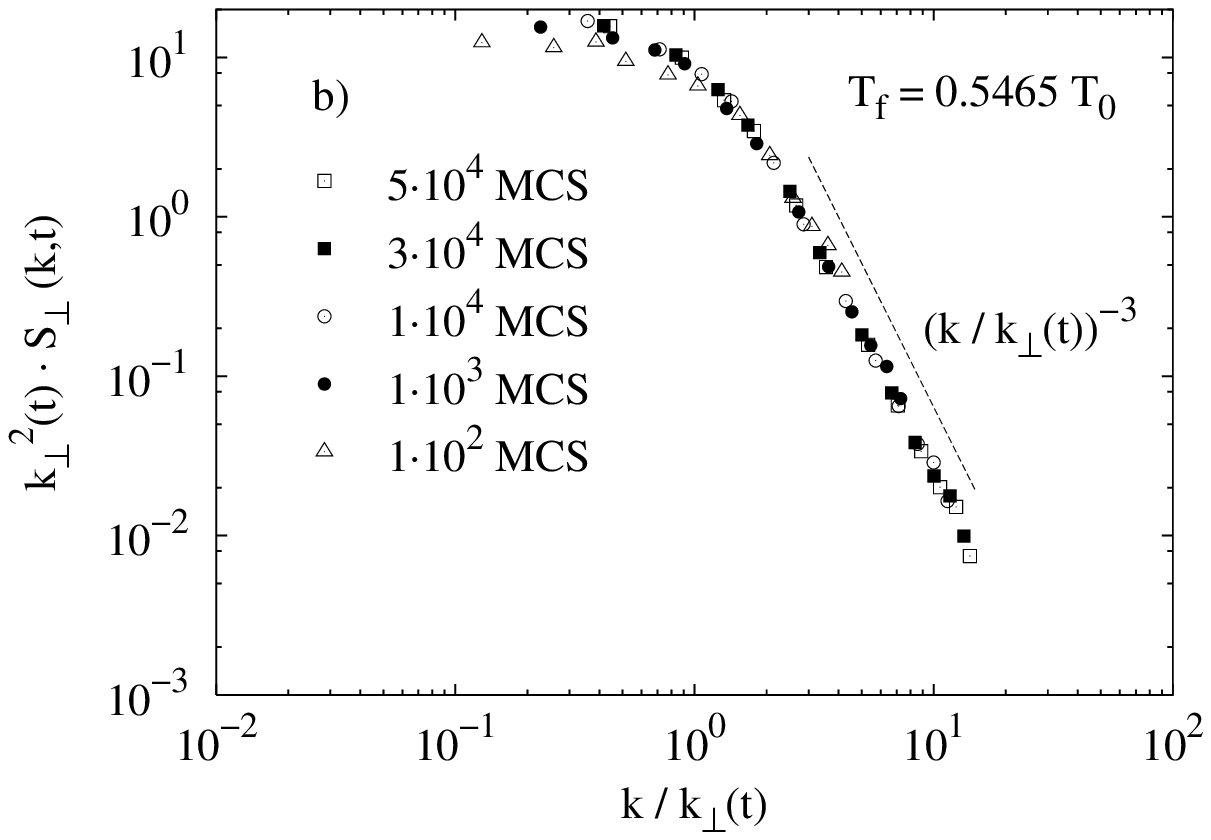, width=8cm}
\caption{Scaled structure factors $k_{\parallel}S_{\parallel}(k,t)$
(a) and $k_{\perp}^2S_{\perp}(k,t)$ (b) calculated for 
$T_f=0.5465\,T_0<T_{sp}$; the slopes of the dashed straight 
lines represent a decay according to  
$(k/k_{\parallel})^{-2}$ (a), and $(k/k_{\perp})^{-3}$ (b).}
\end{figure}

The last point we like to address concerns the observed 
growth rates of the characteristic domain--size, $l(t)$. Since 
the transition considered here is described by a 
set of non--conserved order parameters $\psi_1$ through 
$\psi_3$, one might expect that ordered domains grow 
according to the Lifshitz--Allen--Cahn law $l(t) \propto t^{\nu}$ 
with $\nu=1/2$.\cite{Lifsh,AllenC} Experiments on Cu$_3$Au
indeed were interpreted in terms of this conventional
growth law.\cite{Shan,Noda,Konishi}
In this context, however, one should be aware of the fact
that systems of the type considered may show 
effective growth exponents during relaxation different  
from the Lifshitz--Allen--Cahn law. In the case of a vacancy 
mechanism effective exponents $\nu>1/2$ in principle 
can arise in models based on a non--conserved order 
parameter. In these studies,\cite{Vives, Fratzl} an increased 
$\nu$ was ascribed to an accumulation of vacancies within the
domain boundaries, leading to an enhancement of the interfacial
dynamics. For the present model we have verified, however, that 
the interaction parameters chosen do not favor segregation of
vacancies in the domain boundaries.\cite{Porta}
Another modification of the Lifshitz--Allen--Cahn
law that can act in the opposite direction results from local   
changes in composition within antiphase boundaries
so that the ordering process gets coupled to a (slow)
conserved order parameter component. 

\begin{figure}
\epsfig{file=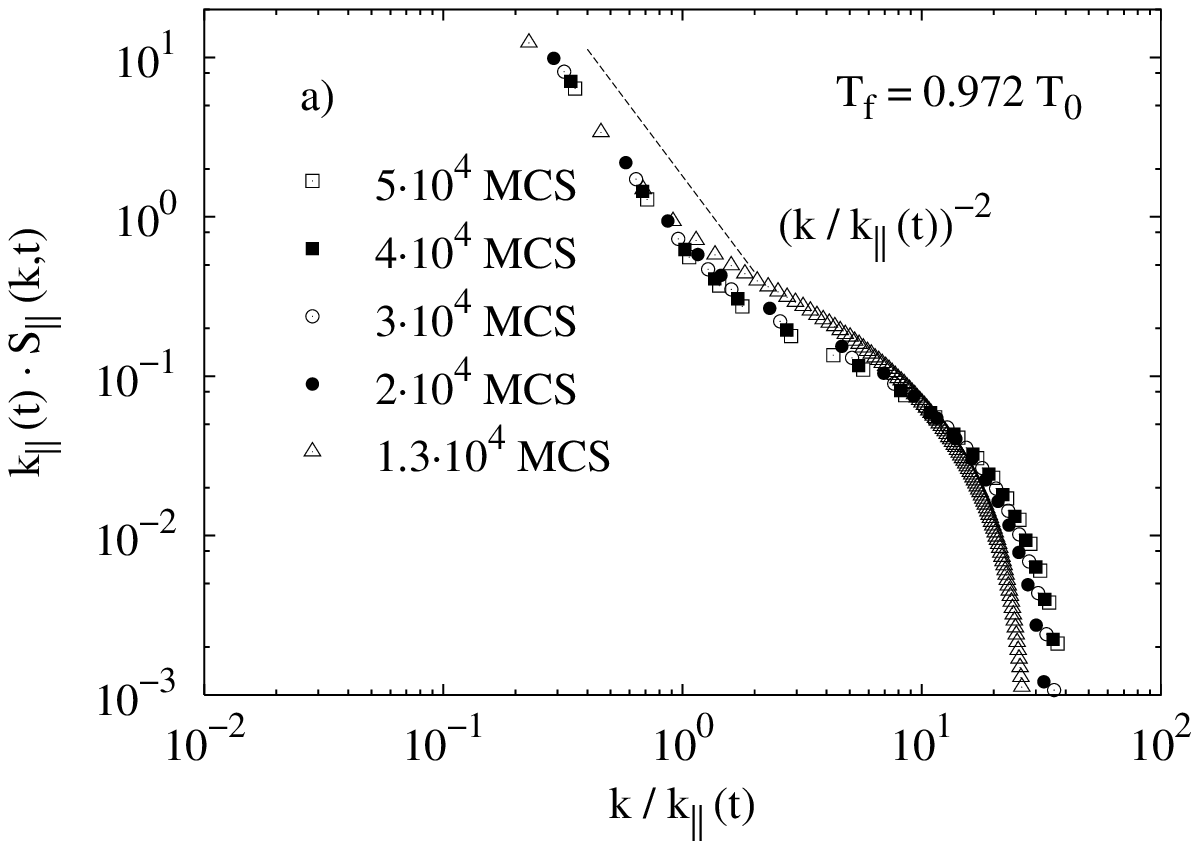, width=8cm}
\epsfig{file=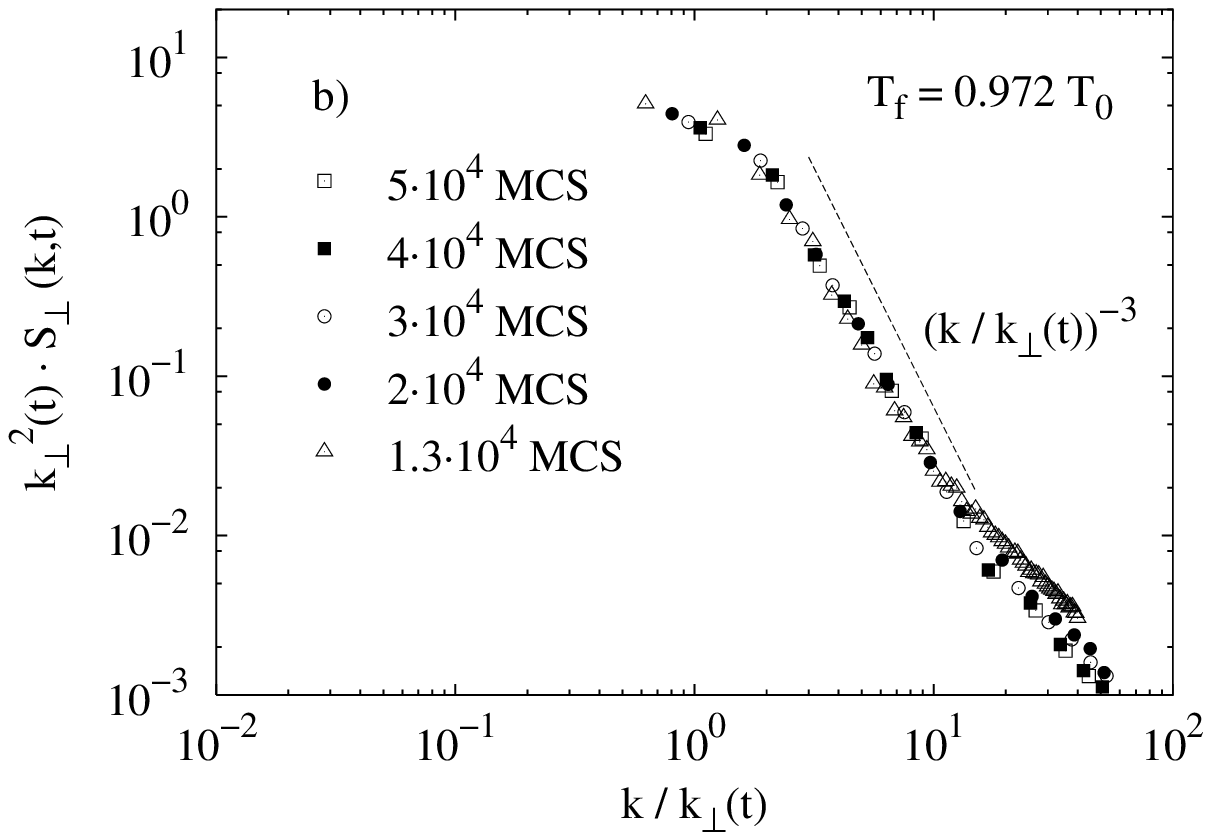, width=8cm}
\caption{Same as in Fig.~7 but for $T_f=0.972\,T_0>T_{sp}$.}
\end{figure}

In our simulations we systematically observe effective growth  
exponents $\nu \simeq 1/4$ or even smaller (cf.~Fig.~5 and 
Kessler {\it et al.} \cite{paper1}). Moreover, within our accessible 
computing times ($3\cdot10^4-~5\cdot10^4$~MCS), these exponents  
depend slightly on temperature. One possible source of these 
differences between simulations and the experimental data 
($\nu \simeq1/2$) seems to be the existence of the low--energy 
type--I walls. Specifically, let us recall that our model
contains only nearest neighbor interactions, so that type--I walls
have zero energy and thus are extremely stable. It has been
suggested previously that such a situation may lead to 
$\nu=1/4$.\cite{Castan,Deymier} Moreover, the present model does 
imply a coupling of the non--conserved order parameters
to the conserved density $\psi_0$. This coupling becomes active within
type--II walls. The relaxation of a modified composition within 
type--II walls is slowed down further because type--II walls are 
interconnected via type--I walls which have fixed $\psi_0$ and do
not allow any exchange of composition.
The slight upward bending of the low--temperature data in Fig.~5
at the longest times, indicating an even slower growth, might
be interpreted in this way, although this point needs 
to be clarified in further studies.

\section{Summary and conclusions \label{SUMMARY}}

Implementing the vacancy mechanism in a model for the atomic 
dynamics in $A_3B$--type fcc--alloys, we investigated the growth
of ordered domains after a temperature quench below the transition 
temperature $T_0$. Depending on the depth of the quench  
we observe two clearly distinct ordering--scenarios: 
thermally activated nucleation of the ordered phase for shallow 
quenches, $T_0>T_f>T_{sp}$, and spinodal ordering, when $T_f<T_{sp}$. 
Here, the spinodal temperature $T_{sp}$ was taken over from independent
simulations at equilibrium.

In the case of thermally activated processes there is some 
characteristic incubation time, $\tau_{inc}$, after which a small 
fraction of the system is covered by overcritical nuclei. Detailed 
simulation results for $\tau_{inc}$, based on vacancy--atom exchange, 
were presented. The time period $t<\tau_{inc}$ manifests itself in 
plateau regions for energy relaxation and for the size of ordered
domains, when plotted versus $t$.
Clearly, $\tau_{inc}$ is expected to diverge as $T_f$ approaches 
$T_0$ from below. Within the simple model investigated here and the 
available maximum computing time,
$\tau_{inc}$ grows more than 260 times in a narrow temperature interval  
above $T_{sp}$ -- from about 90~MCS at $T_f=0.911\,T_0$ to roughly 
$24\cdot10^3$MCS at $T_f=0.976\,T_0$. 
In comparison with measurements \cite{Noda} of $\tau_{inc}$
this seems to constitute the correct order of magnitude
when Monte Carlo times are converted to physical times
with the help of experimental tracer diffusion coefficients. 
In the coarsening regime we observed growth of the
characteristic domain--size that was clearly slower than the 
conventional $t^{1/2}$--law expected for curvature driven 
processes in the presence of non--conserved order
parameters. This may be due to the enhanced stability of the
low--energy domain walls in the case of nearest--neighbor
effective interactions assumed in our model. The enhanced
stability of low energy walls leads to strong anisotropies
of the domain--shapes observed in our system, and furthermore, to
independent scaling laws for the correlation functions
$S_{\parallel}(k,t)$ and $S_{\perp}(k,t)$. These two functions
scale according to Porod's laws for dimensions one 
and two, respectively, when $T_f\ll T_{sp}$. For $T_f>T_{sp}$ 
strong deviations from these scaling laws are observed in the 
region of large $k$--values, due to highly structured 
domain walls which originate from the stage of fast growth 
of ordered nuclei against disordered bulk.

\section*{Acknowledgments}

Helpful discussions with P.~Maass are gratefully acknowledged. 
This work was supported in part by the Deutsche 
Forschungsgemeinschaft, SFB~513.



\begin{thebibliography}{99}
\bibitem{degeneracy}
As a result, the ground state of that model in fact is infinitely
degenerate.
\end{thebibliography}
\end{document}